\documentclass{article}

\usepackage{arxiv}
\usepackage{subfig}
\usepackage{tabularx}
\usepackage[linesnumbered,ruled,vlined]{algorithm2e}
\usepackage[utf8]{inputenc} 
\usepackage[T1]{fontenc}    
\usepackage{hyperref}       
\usepackage{url}            
\usepackage{booktabs}       
\usepackage{amsfonts}       
\usepackage{nicefrac}       
\usepackage{microtype}      
\usepackage{lipsum}		
\usepackage{graphicx}
\usepackage{doi}
\usepackage[export]{adjustbox}

\title{Ensemble of Opinion Dynamics Models to Understand the Role of the Undecided in the Vaccination Debate}

\author{ \href{https://orcid.org/0000-0000-0000-0000}{\includegraphics[scale=0.06]{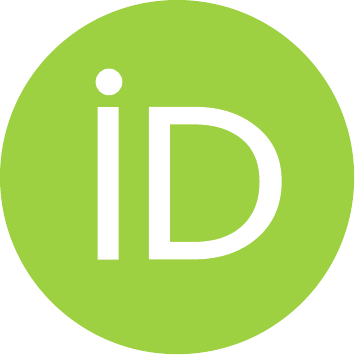}\hspace{1mm}Jacopo Lenti}\thanks{MSc. Thesis Stochastics and Data Science, University of Torino} \\
	ISI Foundation\\
	Torino, Italy \\
	\texttt{jacopo.lenti@isi.it} \\
	\And
	\href{https://orcid.org/0000-0000-0000-0000}{\includegraphics[scale=0.06]{orcid.pdf}\hspace{1mm}Giancarlo Ruffo} \\
	Department of Computer Science\\
	Università di Torino\\
	Torino, Italy \\
	\texttt{giancarlo.ruffo@unito.it} \\
}

\renewcommand{\shorttitle}{\textit{arXiv} Template}

\hypersetup{
pdftitle={A template for the arxiv style},
pdfsubject={q-bio.NC, q-bio.QM},
pdfauthor={David S.~Hippocampus, Elias D.~Striatum},
pdfkeywords={First keyword, Second keyword, More},
}

\begin{document}
\maketitle

\begin{abstract}
In the last years, vaccines debate has attracted the attention of all the social media, with an outstanding increase during COVID-19 vaccinations campaigns. The topic has created at least two opposing factions, pro- and anti-vaccines, that have conflicting and incompatible narratives. However, a not negligible fraction of the population has an unclear position, as many citizens feel confused by the vast amount of information coming from both sides in the online social network. The engagement of the undecided population by the two parties has a key-role in the success of the vaccination campaigns.

In this paper, we present three models used to describe the recruitment of the undecided population by pro-vax and no-vax factions in a three-states context.
Starting from real-world data of Facebook pages previous labelled as pro-, anti-vaccines or neutral, we describe and compare three opinion dynamics models that catch different behaviours of the undecided population.

The first one is a variation of the SIS model, where undecided position is considered an indifferent position, including users not interested in the discussion. Neutrals can be ``infected'' by one of the two extreme factions, joining their side, and they ``recover'' when they lose interest in the debate and go back to neutrality.

The second model studied is a Voters model with three parties: neutral pages represent a centrist position. They lean their original ideas, that are different from both the other parties. 

The last is the Bilingual model adapted to the vaccination debate: it describes a context where neutral individuals are in agreement with both pro-, ad anti-vax factions, with a position of compromise between the extremes (``bilingualism''). If they have a one-sided neighbourhood, the necessity (or the convenience) to agree with both parties comes out, and bi-linguists can become mono-linguists.

Our results depicts an agreement between the three models: anti-vax opinion propagates more than pro-vax, thanks to an initial strategic position in the online social network (even if they start with a smaller population). While most of the pro-vaccines nodes are segregated in their own communities, no-vaccines ones are entangled at the core of the network, where the majority of undecided population is located.

In the last section, we propose and compare some policies that could be applied on the network to prevent anti-vax overcome: they lead us to conclude that censoring strategies are not effective, as well as segregating scenarios based on unfollowing decisions, while the addition of links in the network favours the containment of the pro-vax domain, reducing the distance between pro-vaxxers and undecided population. 
\end{abstract}



\section*{Introduction}
Far before COVID-19 pandemics, no-vax messages have spread all over the world. In the last decade, vaccinations rates have decreased in many countries with severe consequences: while in 2000 measles was declared eliminated from US by the WHO, in recent years the reported cases had an outstanding growth, reaching 1215 cases in 2019; numerous outbreaks of measles plagued many places with low vaccinations rates: in 2019, 83 measles deaths  in unvaccinated population were reported in Western Samoa, 338 in Philippines; in the countries of the WHO European Region, cases of measles leapt from 5,273 in 2016, to 83,540 in 2018; in 2015 it was recorded the first death of diphtheria in Spain after 28 years, on a non-vaccinated child; the vaccine hesitancy has caused the resurge of polio in 20 African countries, previously considered polio-free. In 2019, Word Health Organization (WHO) named vaccine hesitancy as one of the top 10 threats to global health.\cite{evencovid} \cite{detecting} \cite{measlesdata1} \cite{measlesdata2}\\
Despite the great efforts and the unprecedented success related to coronavirus vaccines discovery and delivery, the risk that the vaccine hesitant population could undermine vaccination campaign is really concrete. As regards coronavirus, in 2021 November, 65\% of US population is fully vaccinated, Mexico 58\%, Russia 41\%, still very far from herd immunity. \cite{databasecovid}\vspace{1pt}

It is important to differentiate vaccine refusal from vaccine hesitancy. Vaccine refusal is a clear position rejecting any vaccines, often for political or ethical reasons. No-vaxxers, often incited by false information, sustain strong and polarised ideas, with an anti-scientific inclination. Their arguments are the same against all the vaccines: the benefits of vaccines are overwhelmed by the side effects, mandatory vaccinations infringe on civil rights and religious beliefs. The anti-vax population refuses dialogue with scientific views and are very resistant to change.\\ 
Despite the outsized attention given to anti-vaccinism by mainstream media, the vast majority of individuals avoiding vaccines can't be considered completely no-vax. Vaccine hesitancy comprises a wide range of ideas between vaccines refusal and vaccines trust. The fraction of vaccines hesitant population is driven by more complex dynamics and middle-ground positions. Most of them are not against vaccines: they can be not confident of some specific vaccines; their positions can be related to feelings of anxiety towards medical treatments or distrust on healthcare system and pharmaceutical companies. Compared to anti-vaxxers, vaccines hesitant individuals support much less extreme arguments, and they can change mind depending on the context. \cite{efforts} \cite{pinterest} \cite{pandora}\cite{trust}\cite{factor}\cite{understanding} \\
For this reason, the efforts of the scientific community and healthcare system should address the vaccine hesitant fraction of population, battling the risk of exposing them to misleading anti-vaccines narratives.\\
In the Internet era, the Web plays a key role in any kind of information spread. News can propagate in social media without filters, and people are hourly exposed to any kind of content, facing difficulties in discriminating high and low-credible sources. Having access to any type of information has been suggested as a potential influence on the growth of anti-vaccination groups. Social media seems to be a powerful promoter of different sentiments about vaccination, hence it is likely that it contributes to vaccines avoidance. \\
Fake news can spread faster than true news, with wider and deeper cascades. A variety of fake news and conspiracy theories related to vaccines polluted the social media, from the famous fraudulent study relating vaccines and autism, published in 1997 by Andrew Wakefiled, to the theory of the worldwide genocide planned by Bill Gates with coronavirus vaccines \cite{vosoughi18} \cite{vaccineautism}\cite{socmediavachesitancy}. \\
Vaccination hesitancy and susceptibility to misinformation are linked. Studies show that parents who exempt children from vaccination are more likely to have obtained information from the Internet than parents who have their children vaccinated~\cite{susceptibility}. \\
Another significant aspect is the linkage between access to information via Web and echo chambers. As a matter of fact, social media tend to suggest the following and interacting with profiles and pages that share similar contents, have similar views and agree with the users. Users select information adhering to their system of beliefs and tend to ignore dissenting information, forming the so-called echo chambers, polarised groups of like-minded people who keep framing and reinforcing a shared narrative. For example, Facebook is often claimed to accelerate the emergence of echo chambers, in which pro- and anti-vaccination attitudes are constantly reinforced, bringing a worrying polarisation of the users \cite{polarizationandfakenews} \cite{polarization}. The existence of echo chambers may explain why social-media campaigns, that provide accurate information, have limited reach and be effective only in sub-groups, even fomenting further opinion polarisation. Furthermore, when polarisation is high, misinformation might easily proliferate~\cite{polarizationandfakenews} \cite{polarization}.\vspace{1pt}

The main purpose of our paper is to extend the study presented by Johnson et al. (2020) \cite{competition}: they considered a network of Facebook pages involved in the vaccination debate, labelled as pro-vaccines, anti-vaccines or neutral. In their study, an anti-vaccines overcome is predicted in few years. We adopt a comprehensive approach, studying the effects of opinion dynamics on the static network, modelling the behaviour of the undecided population in a news spread. As an output of our study, we want also to suggest some strategies that is more likely to succeed to restrain anti-vaxxers convincing the undecided.\\
In general, an idea or an opinion (such as views in favour or against a vaccine) is often transmitted as an intentional act from one person to  another. Through distinct types of interactions, people want to persuade others to adopt an  opinion. The views of others might have an impact on individual beliefs, who update their own opinion. Different ways to model the underlying interactions and the updating process are studied.\\
Opinion dynamics has been modelled through a variety of angles and techniques, for instance, kinetic models of opinion  formation, mean-field analysis, agent-based models and epidemiological  models. Individual opinion can be a number in an interval, such as [-1,1], where the two extremes represent the two opposing opinions; on the other hand, it can be a categorical variable, where each agent can be assigned to one of the few possible compartments (usually two).
According to the position with respect to the two opinions, and through interactions, individuals move their position, possibly influencing others: many elements can be employed in the model to drive the phenomenon, such as memory loss, the presence of leaders, the ability to convince others with a different opinion, varying levels of assertiveness, the fact that more extreme opinions are more difficult to change. \cite{opinionepidem} \cite{opinionpersuasion} \cite{opinionleader} \cite{opinionstats} \cite{opinionsegr} \cite{opinioncrime} \cite{opinionkine}

\section*{Data}
\begin{figure}[!htb]
    \centering
    \includegraphics[width = 0.8\textwidth]{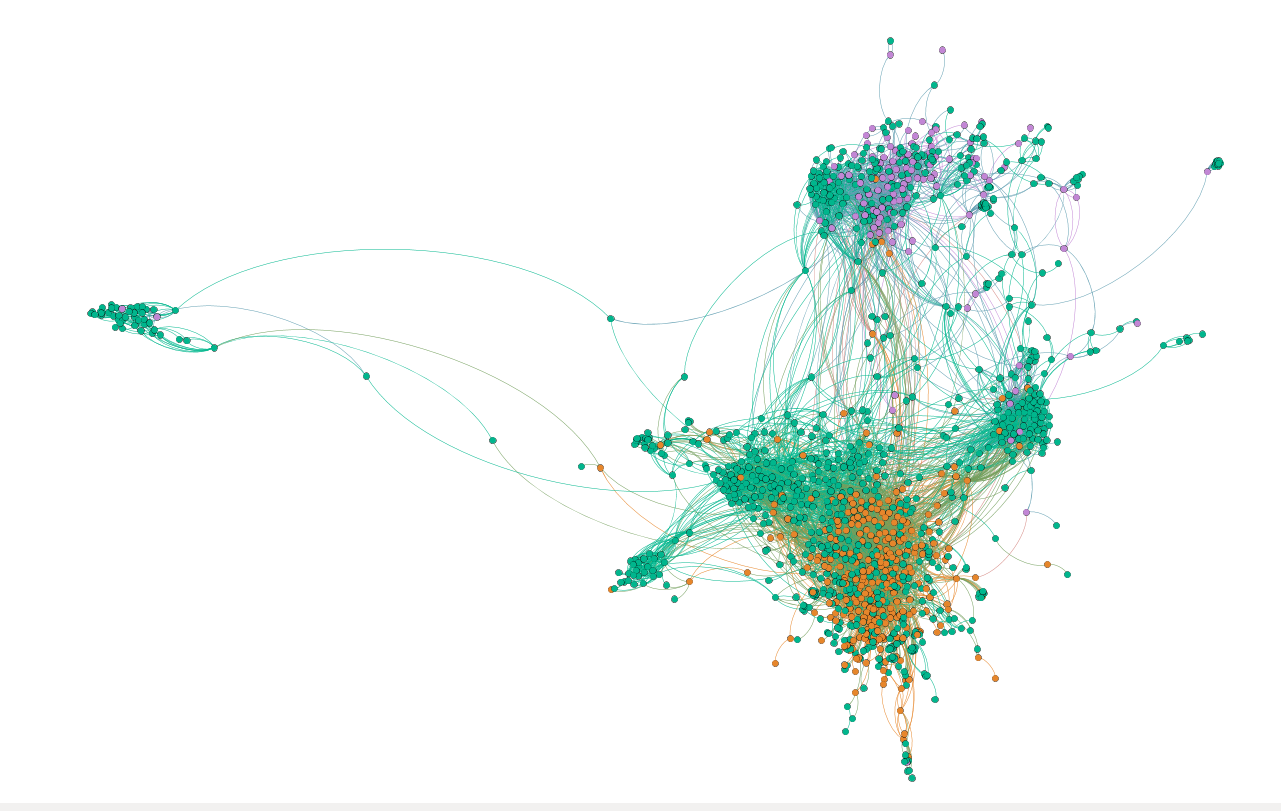}
    \caption{Graphical representation of the social network studied. Orange nodes are anti-vaccines pages, blue pro-vaccines, and green neutral. It is possible to note a community structure: one large community with most of the network population is located at the bottom of this figure (mostly green and orange); the second most populated community, provax, at the top; two smaller communities, one on the left and one on the right, neutrals with a few differently coloured nodes.}
    \label{fig:my_label}
\end{figure}
The dataset used in this study has been introduced by Johnson et al. (2020) \cite{competition}. Each node of the network is a Facebook page, and edges represent a page recommending (i.e. ``fanning'') another fan page. Therefore, a Facebook user can see the out-link as a recommendation from an authority that he or she supports. The data are collected in 2019: for each page we know its connections to other pages, its popularity (the number of followers), and the faction it supports (pro-vaccines, anti-vaccines or neutral). A neutral page is a page that is involved in the debate around vaccination, but has not expressed a clear side.\\
The network is made of 1326 pages (nodes) and 7484 links. Most of the nodes belongs to neutral party (66.7$\%$), followed by no-vaccines (23.9$\%$) and pro-vaccines (9.4$\%$). However, since no-vax pages are much smaller, total pro-vax population is larger than no-vax one, with 7 millions users versus 4 millions, while 74 millions users are neutral.\\
It is possible to observe that pro-vax nodes tend to remain at the periphery of the network, in their tightly-knit communities. On the other hand, most of no-vax nodes are sparse at the core of the network, with many small pages connected to many central neutral clusters.
\begin{table}[!htb]
    \centering
    \begin{tabular}{|c|cc|cc|c|}
    \hline
         & \#Nodes &  & \#Fans & & \#Fans per page \\
    \hline
    Pro-vaccines & 124 & 9.3\% & 6.9M & 8.1\% & 55.1k\\
    Neutral & 885 & 66.7\% & 74.1M & 87\% & 83.9k\\
    Anti-vaccines & 317 & 23.9\% & 4.2M & 4.9\% & 13.1k\\
    \hline
    \end{tabular}
    \caption{Summary table of the proportion of nodes and fans per faction.}
    \label{tab:my_label}
\end{table}

\section*{Methodology}
In this section we report the ensemble of opinion dynamics models that we have studied. The models describe the behaviour of neutral pages during the discussion around a piece of news related to vaccination debate. 
At first, we simulate the news spreading: the nodes share the piece of news to their neighbourhood and are involved in the debate. Secondly, we model opinion dynamics, and pages aware of the news can be influenced by or can influence some neighbours.\\
We assume that nodes that are initially pro-vax or no-vax will remain in that state forever, while an undecided vertex can change state many times. Direct transitions from one extreme to the other are not allowed.\\
We can compare the proportion of pages belonging to the factions after 400 iterations, analysing the analogies and differences between the three models. We want to understand and explain the effects of the different hypotheses formulated.\\


\paragraph*{Definitions}
\begin{itemize}
    \item Nodes: the nodes of the social network (also \emph{vertices}, or \emph{pages}). They can be anti-vaccines, pro-vaccines or neutral. For each node we know the initial state and the current state at each step. Moreover, nodes are either \emph{aware} or \emph{ignorant} about the piece of news discussed.
    \item $\gamma$: transmission rate of news spread.
    \item \emph{d}: dampling factor. It is the probability that a random page become \emph{aware} of the news from an external source.
    \item $\beta_A$, $\beta_P$: the rate at which anti-vax (\emph{A}) or pro-vax (\emph{P}) nodes influence neutral neighbours.
    \item $\mu_A$, $\mu_P$: the rate at which originally neutral pages go back to neutrality from anti-vax or pro-vax party.
    \item $\sigma^i_K$: the proportion of neighbours of node \emph{i} in the faction \emph{K} (with \emph{K} in $\{ A, P, N \}$).
    \item $p^i_{K_1 \leftarrow K_2}$: transition probability of node \emph{i} from $K_1$ to $K_2$.
\end{itemize}

\subsection*{News spread}
Nodes can be \emph{ignorant} or \emph{aware} of the news. Following epidemic framework, an SI Model guides the news spread: \emph{aware} nodes are considered as Infectious, and \emph{ignorant} are Susceptible. An \emph{aware} page can share the piece of news to some \emph{ignorant} neighbours, making them changing state to \emph{aware}. An \emph{aware} vertex will remain \emph{aware} forever; hence, the only admitted transition is from \emph{ignorant} to \emph{aware}.\\
At each step, a page \emph{i} picks one of its neighbours, \emph{j}, with probability proportional to their popularity. If the page \emph{i} is \emph{ignorant} and \emph{j} is \emph{aware}, with probability $\gamma$ the \emph{ignorant} page \emph{i} is involved and become \emph{aware}.\\
Moreover, it is possible that a node becomes \emph{aware} from some source coming from outside of the network: at each iteration, with probability \emph{d} (dampling factor), a random page becomes \emph{aware}.
\subsection*{Polarisation Model}
Three models are compared to predict pro-vax and no-vax influence on neutral party. They share some common features: at each step, an initially neutral page selects one of its neighbours with probability proportional to their popularity; subsequently, the extracted page can influence the previous one (following transition probabilities that depend on the model). In 80$\%$ of cases, nodes select their out-neighbours, in 20 $\%$ they select their in-neighbours. This assumption depicts the higher probability to be influenced by out-neighbours, without bringing constraints related to strongly connectivity.
\paragraph{SIS Model}
The first model we present is an adaptation of the SIS Model. It is a well-known compartmental epidemic model, widely used in opinion dynamics and social networks. The population is split into Susceptible and Infectious: Susceptible individuals can be infected by some neighbours and move to Infectious state; after a certain amount of time, an Infectious individual can recover and turn to Susceptible state. Thus, a recovered individual does not become immune, and can be infected again.\\
Here, we consider neutral nodes as Susceptible, while pro- and no-vax nodes represent two different kinds of Infectious actors, that cannot infect each other, although they can transmit their own `disease' (i.e., opinion) to neutrals. At each step, Infectious pages (originally neutral) can recover and become neutral. \\
This model depicts social contagion, where undecided pages can be considered as external to the debate: after some time pages can lose interest and move spontaneously from one extreme to neutrality. Neutral side represents an indifferent position.\\
We define the following parameters: $\beta_A$ and $\beta_P$ the infection rate for anti-vaccines and pro-vaccines contagion, $\mu_A$ and $\mu_P$ the recovery rate from anti-vaccines and pro-vaccines infection. Letting $\sigma_A^i$ and $\sigma_P^i$ be the proportion of anti-vax and pro-vax neighbors of node \emph{i}, we obtain these transition probabilities for node \emph{i}:
\begin{itemize}
\item $p^i_{N \rightarrow P}$: $\sigma^{\textit{i}}_{P} \beta_{P}$;
\item $p^i_{N \rightarrow A}$: $\sigma^{\textit{i}}_{A} \beta_{A}$;
\item $p^i_{P \rightarrow N}$: $\mu_{P}$;
\item $p^i_{A \rightarrow N}$: $\mu_{A}$.
\end{itemize}

\paragraph{Voters Model}
The second model analysed is an extension of the Voters Model, where a third option is introduced. The original Voters Model describes an interacting particle system, where each voter can be influenced by one of its neighbours to change opinion. Following previous literature \cite{freezing}, a ``centrist'' party is introduced, and we assume incompatibility between pro- and no-vax view. All the transition probabilities have the same form: neutral clusters can persuade new users with a different opinion. Neutrality is not an indifferent position, but it is the support of a different narrative. \\
Let us define the parameters used: $\beta_A$ and $\beta_P$ are the persuasion rates from neutrality to anti-vaccines and pro-vaccines parties; $\mu_A$ and $\mu_P$ are the persuasion rates from the extremes to neutral centrist party.\\
The resulting transition probabilities for a node \emph{i} are the following:
\begin{itemize}
\item $p_{A \rightarrow N}$: $\beta_{A} \sigma^{\textit{i}}_{N}$;
\item $p_{P \rightarrow N}$: $\beta_{P} \sigma^{\textit{i}}_{N}$;
\item $p_{N \rightarrow A}$: $\mu_{A} \sigma^{\textit{i}}_{A}$;
\item $p_{N \rightarrow P}$: $\mu_{A} \sigma^{\textit{i}}_{P}$.
\end{itemize}
It is proved that Voters Model will always reach an equilibrium, where no transitions are allowed: it can be a consensus equilibrium (all the nodes have the same polarity), or a frozen mixture of the two extremes (if no neutral nodes are present, it is impossible to have new transitions) \cite{ultimatefate}. Since some nodes are pro- or no-vax forever by assumption, the only possible fate is to obtain a neutral-free population.

\paragraph{Bilingual Model}
The third model in consideration is Bilingual Model, coming from the language competition. Here, we consider neutral nodes as bilinguist individual, that are able to communicate with both the parties. A monolinguist (pro-vax or no-vax node in our context) can decide to adopt the opposing language and become bilinguist if it has the incentive to communicate with some of its neighbours. A bilinguist can be induced to abandon one of the two languages if it has not the necessity to know it to speak to its friends \cite{orderingbilingual}.\\
This model can be applied to those news where we can assume that a part of the users can agree with both the opposing factions. Bilingualism is a compromise solution.\\
Let us denote as $\beta_A$ and $\beta_P$ the incentives to monolinguism for neutral pages, and $\mu_A$ and $\mu_P$ the incentive to bilinguism for the extreme nodes. The assumptions lead to the following transition probabilities:
\begin{itemize}
\item $p^i_{A \rightarrow N}$: $\sigma^{\textit{i}}_{P} \beta_{A}$;
\item $p^i_{P \rightarrow N}$: $\sigma^{\textit{i}}_{A} \beta_{P}$;
\item $p^i_{N \rightarrow A}$: $(1 - \sigma^{\textit{i}}_{P}) \mu_{A}$;
\item $p^i_{N \rightarrow P}$: $(1 - \sigma^{\textit{i}}_{A}) \mu_{P}$;
\end{itemize}
Notice that $1 - \sigma^{\textit{i}}_{P}$ is equal to $\sigma^{\textit{i}}_{A}$ + $\sigma^{\textit{i}}_{N}$, so the incentive for a neutral page to move to anti-vaccines side is given by the fraction of neighbour that agree with no-vax position, that is the sum of no-vax and neutral clusters. 

\section*{Algorithm}

\begin{algorithm}[!htb]
\caption{Opinion Dynamics}
\SetKwFunction{RandomAct}{\normalfont NEWS SPREAD RANDOM}

\SetKwProg{Fn}{function}{:}{}
\Fn{\RandomAct}{
Pick a random node $i$\\
With probability $d$ set $i$ \texttt{aware}\\
}

\vspace{5pt}

\SetKwFunction{NeighbourAct}{\normalfont NEWS SPREAD NEIGHBOURS}

\SetKwProg{Fn}{function}{:}{}
\Fn{\NeighbourAct}{
\For{i \emph{in nodes}}{
Pick one random neighbour $j$ \\ 
\If{j \emph{\texttt{aware}}}{With probability $\gamma$, set $i$ \texttt{aware}}
}
}
\vspace{5pt}

\emph{t} = 0 \\
Set all nodes \texttt{ignorant}\\
Set one random node \texttt{aware}\\
\While{t \upshape < total time}{
$t$ = $t$ + 1\\
NEWS SPREAD NEIGHBOURS\\
NEWS SPREAD RANDOM\\
\For{i \emph{in nodes initial \texttt{neutral}}}{
Select a random neighbour $j$ with probability proportional to \texttt{fan\textunderscore count}.\\
\eIf{i \emph{\texttt{neutral} \& \emph{j} \texttt{not} \texttt{neutral}}}{
$i$ can assume the state of $j$ under the rules of the model.\\}{
\If{i \emph{\texttt{not neutral}}}{
set $i$ \texttt{neutral} under the rules of the model
}

}
}

}

\end{algorithm}

\section*{Results}
After the description of the models, let us present the results obtained by simulating their behaviours in the network.\\
We repeated the simulations by varying the four parameters for each of the models:
\begin{itemize}
    \item \emph{d} = 0, 0.2;
    \item $\gamma$ = 0.025, 0.05, 0.1, 0.2;
    \item $\beta_A$ = $\beta_P$ = $\beta$ = 0.1, 0.2, ..., 0.9;
    \item $\mu_A$ = $\mu_P$ = $\mu$ = 0.1, 0.2, ..., $\beta$.
\end{itemize}
For all the possible configurations of parameters two simulations were performed: one starting the news spread from a no-vax node, and one from a pro-vax. For each model 720 simulations were carried out, one for each possible vector of parameters. The simulations stop when an equilibrium is reached. \\
Under the conditions $\beta_A$ = $\beta_P$ and $\mu_A$ = $\mu_P$, anti- and pro-vaccines parties have the strength, both in persuading new users and in losing them.\\
Notice that $\beta$ $\geq$ $\mu$ in all the cases: in general, we are assuming that extreme options are more attractive than neutral option.\\

\begin{figure}[!htb]
\centering
\def\tabularxcolumn#1{m{#1}}
%

\begin{tabular}{cc}

\subfloat[\emph{t =} 0]{\includegraphics[width=4cm]{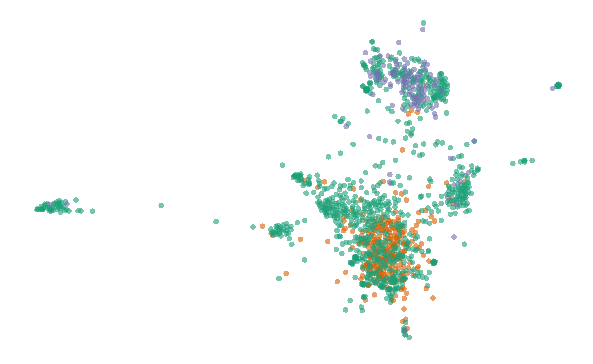}} &
\subfloat[\emph{t =} 5]{\includegraphics[width=4cm]{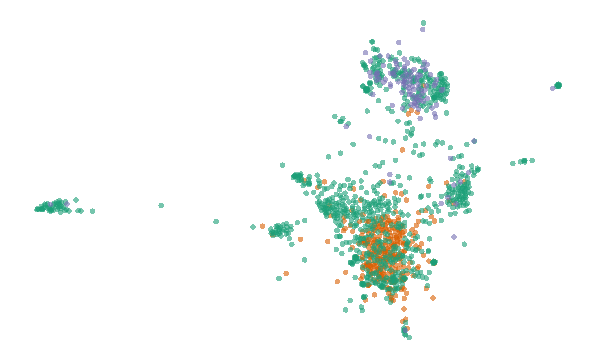}} \\
\subfloat[\emph{t =} 20]{\includegraphics[width=4cm]{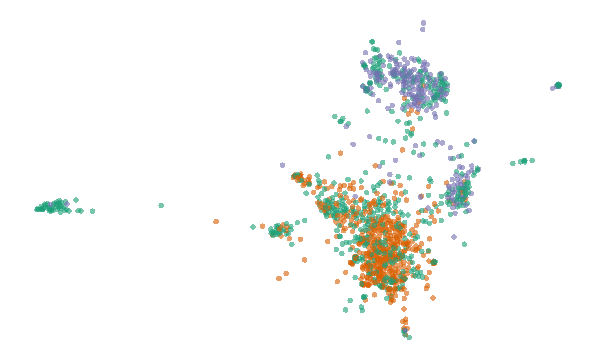}} &
\subfloat[\emph{t =} 100]{\includegraphics[width=4cm]{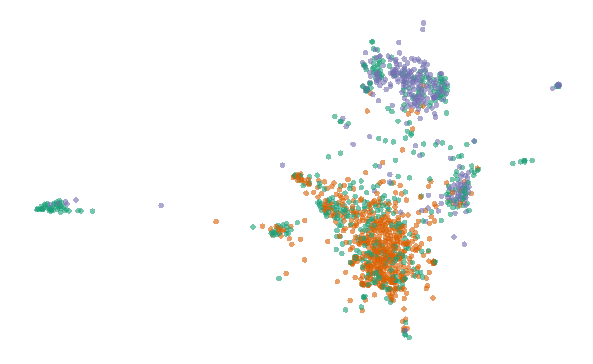}} 
\end{tabular}


\caption{\emph{SIS Model} Graphical representation of a simulation, where $\mu$ = 0.2, $\beta$ = 0.4, $\gamma$ = 0.1 and \emph{d} = 0.2, with snapshots at different times.\\
Blue nodes represent pro-vax pages, green neutral and orange no-vax.\\
Since $\mu < \beta$, the number of neutral pages decreases, settling around 500-550. Each community assumes the colour that was more present at the beginning, with many green points placed at random.}
\label{SIS_network}

\end{figure}

\paragraph{SIS Model}
In Fig. ~\ref{SIS_network} a simulation example is presented. In agreement with epidemic SIS model, the number of neutral/susceptible nodes tends to oscillate around an equilibrium value, hence, a consistent fraction of neutral pages survives in all the simulations. Since the recovery does not depend on the surrounding context, neutral clusters are located completely random. The two factions grow nearly in the same way as number of pages. However, no-vax community grows much more as number of users, since they start with few fans and they attract many popular clusters. In the first phases there is no competition at the core of the network, so no-vax party engages many neutral clusters, even if they are not much popular. Pro-vax pages start to persuade clusters in their communities, but they remain too far from the centre of the network.\\

\begin{figure}[!htb]
\centering
\includegraphics[width = 0.4\textwidth]{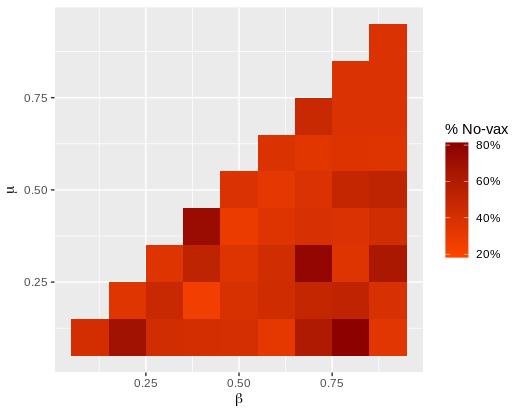}
\caption{ Heatmap representing the average percentage of no-vax users for each pair of $\mu$ and $\gamma$.}
\label{SIS_heatmap}
\end{figure}
While the number of pro-vaccination pages is always smaller than the number of anti-vaccination ones, in 75$\%$ of simulations the sum of the followers of pro-vax side pages remains larger than the sum of no-vax side followers. On average, anti-vaccinist view reaches 44$\%$ of non-neutral pages, not far from the tie (starting from 37.5$\%$).\\
Fig. \ref{SIS_heatmap} shows the effect of the variation of $\beta$ and $\mu$. No-vax cluster takes advantage of low values of $\mu$, independently of $\beta$. When it is unlikely to lose interest around a topic, the initial disposition of anti-vaccines clusters is more effective, because neutral pages follow the first opinion heard and maintain it during the time. Discussions that resurface many times tend to favour the seeking of news from many sources, making popular pages more influent.\\
There is an advantage in being the first spreader of the news: having the initial seed, anti-vaccines community wins the competition in 27$\%$ cases, that decreases to 22$\%$ if the starter is pro-vaccinist.

\paragraph{Voters model}
In Voters Model with centrist option, an equilibrium is always achieved: since a part of the nodes are set to be pro-vax or no-vax forever by assumption, the only possible equilibrium is the mixture of the two extremes. Under this configuration, the network is split in two factions that can't communicate each other and do not seek for a compromise.\\

\begin{figure}[!htb]
\centering
\def\tabularxcolumn#1{m{#1}}
%

\begin{tabular}{cc}

\subfloat[\emph{t =} 0]{\includegraphics[width=4cm]{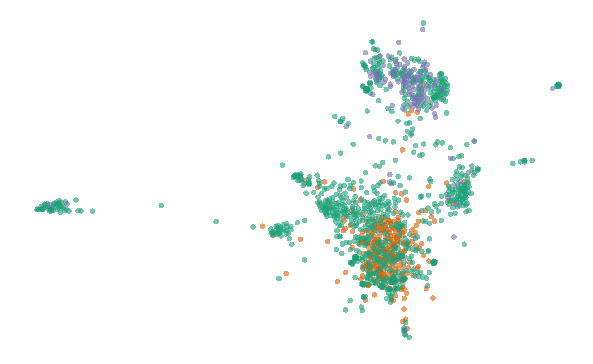}} &
\subfloat[\emph{t =} 5]{\includegraphics[width=4cm]{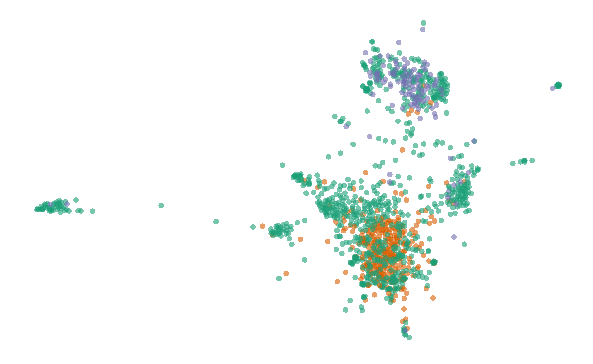}} \\
\subfloat[\emph{t =} 20]{\includegraphics[width=4cm]{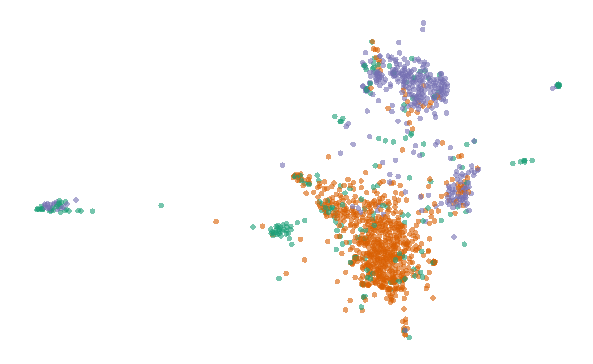}} &
\subfloat[\emph{t =} 100]{\includegraphics[width=4cm]{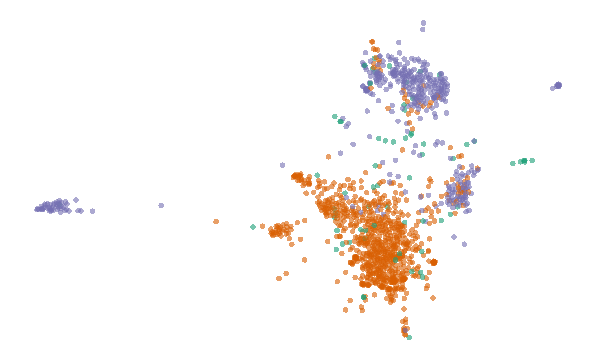}} 
\end{tabular}
\caption{\emph{Voters Model}  As before, the parameters of the simulation are $\mu$ = 0.2, $\beta$ = 0.4, $\gamma$ = 0.1 and \emph{d} = 0.2.
}
\label{VOT_network}
\end{figure}

Looking at the example in Fig. \ref{VOT_network}, we can observe that communities tend to reach the consensus of the initial majority, favouring an high polarisation. The exception is given by the small community on the right, that has half of nodes orange and half blue: since these nodes have to  cohabit with opposing neighbours, this could lead to conflicting conditions, because none of them is prone to change opinion or to seek a compromise.\\
It is unlikely that a page engaged during the initial stages decides to change side in a later moment: since the transitions depend on the neighbourhood, this can happen only when some other neighbours change faction.\\
The pro-vax victories are obtained in only 31$\%$ of cases. No-vax pages, that are well infiltrated at the core of the network, are able to attract many neutral users in the initial phases: in this way they can create a orange populated community, that can survive favouring no-vax domain. These dynamics are not present in the SIS model, where at each moment the engaged pages can recover, making less decisive the initial configuration.\\ 

\begin{figure}[!htb]
\centering
\def\tabularxcolumn#1{m{#1}}
%

\begin{tabular}{cc}

\subfloat[\emph{t =} 0]{\includegraphics[width=4cm]{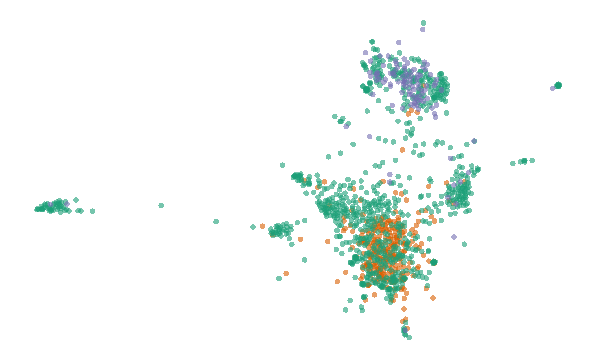}} &
\subfloat[\emph{t =} 5]{\includegraphics[width=4cm]{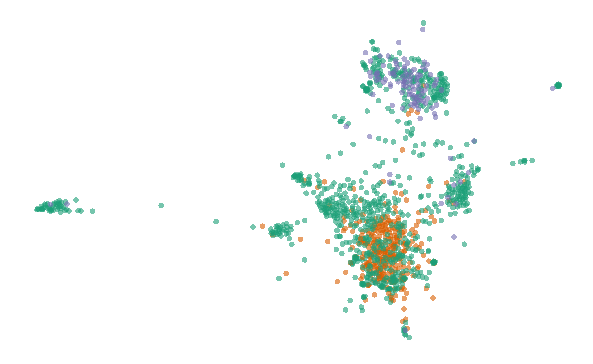}} \\
\subfloat[\emph{t =} 20]{\includegraphics[width=4cm]{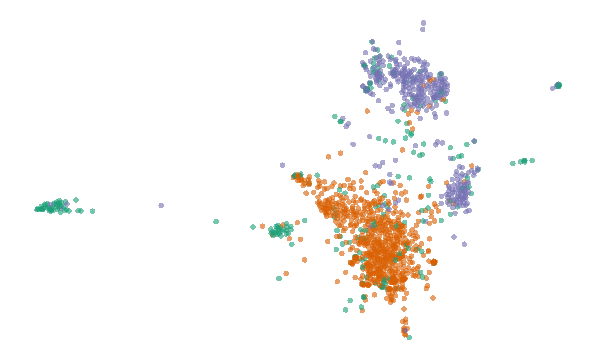}} &
\subfloat[\emph{t =} 100]{\includegraphics[width=4cm]{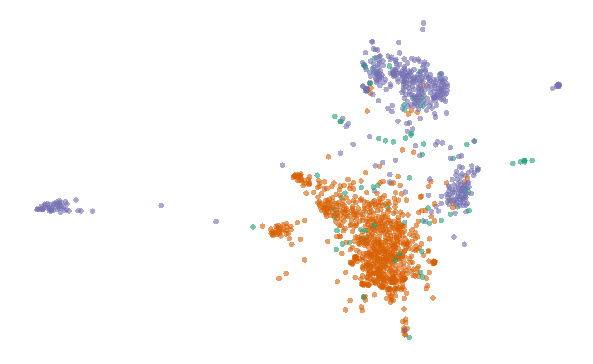}} 
\end{tabular}

\caption{\emph{Bilingual Model} The parameters of the simulation are $\mu$ = 0.2, $\beta$ = 0.4, $\gamma$ = 0.1 and \emph{d} = 0.2.
}
\label{BIL_network}
\end{figure}

\paragraph{Bilingual Model}
Bilingual option is needed only by individuals that have the necessity to communicate to both the parties, while nodes whose neighbours belong to the same party are induced to abandon the less useful language:  for this reason, communities tend to assume homogeneous colouring, and bilinguists are only present at the border of communities, acting the role of buffer states. As we can see from Fig. \ref{BIL_network}, the central community is completely converted to no-vax opinion, and peripheral clusters are under pro-vax influence. With the analogous argument presented in Voters Model, we can highlight that the initial configuration is very influent in the final outcome.\\ Anti-vaccines party wins the competition in 84$\%$ of the simulations, the best performance among the models. Pro-vax opinion is confined in peripheral regions, and the neutral border obstructs the approach to the centre of the network.\\

The simulations tell us that the initial configuration of the network favours no-vax overcome: in Voters and Bilingual Models, no-vaxxers win in most of cases, and in SIS Model they are able to reduce the gap from pro-vax. Since no-vax nodes are widely present around neutral pages, undecided users tend to be attracted by no-vax opinion, just because pro-vax is too far from them. Despite no-vax community is much less populated than pro-vax one, it has a noticeable influence given by their high number of pages placed in strategic locations.\\
A strong polarisation arises in our simulations: after the news spread, it is likely that a node at the centre of a community supports an opinion in agreement with all its neighbourhood. It is not aware of the opposing view, because all the contents it can see have the same polarity. This phenomenon is more observable in Voters and Bilingual cases, where all the pages of the network are involved in the debate, and no one is indifferent. Moreover, it is not possible to have an opinion that differs from the surrounding, because the transitions always depend on the opinion of the neighbourhood (people can't forget or lose interest on the discussion).\\

\subsection*{Pro-vax Strategies}
Having seen the risk related to a no-vax opinion spread, our goal is to study some interventions that could be applied to prevent their overcome, exposing undecided population to pro-vax view.\\
We compare some policies that could be adopted to change the topology of the network by pro-vaxxers: they can be either aimed to widen pro-vaccines audience, or to obstruct no-vax message. In the first case, we create new links targeting pro-vax nodes (meaning that they are increasing their followers). Depending on the nodes involved, different strategies are explored. Secondly, some no-vax nodes or links are removed from the network, miming the behaviour of many social networks that delete misleading contents or pages.\\
To sum up, 6 different strategies are proposed:
\begin{enumerate}
    \item \emph{Random Linkage}: pro-vax pages enlarge their audience to anyone in the same way. New links are added from random pages to pro-vax pages.
    \item \emph{Neutral Linkage}: in our models it is impossible to make anti-vaccines pages change side, so it could be useful to focus on neutral community. New links start from random neutral clusters to pro-vax pages, enlarging only the public of influenceable users.
    \item \emph{Neutral Popular Linkage}: neutral pages are picked with probability proportional to their popularity. Since transmissions depend on the number of fans of the pages, this strategy address influenceable and influent pages.
    \item \emph{Neutral Influent Linkage}: neutral pages are extracted with probability proportional to the number of neighbours. Engaging pages with a high number of neighbours could provide a way to influence more pages.
    \item \emph{Links Removal}: random links starting from neutral pages addressing no-vax pages are removed.
    \item \emph{Nodes Removal}: random anti-vaccines pages are removed from the network.
\end{enumerate}
We studied the effects of these strategies by simulating the diffusion of news after having applied the policies on the networks.\\
The number of links added varies between 0 and 4000 (the number of links grows between 0$\%$ and 53$\%$), while the number of links removed range the interval 0-1833 (remove from 0 and 100\% neutral-novax links), and the number of nodes removed varies between 0 and 317 (remove from 0 to 100 \% of anti-vax nodes).\\
\begin{figure}[!htb]
\centering
\includegraphics[width = \textwidth]{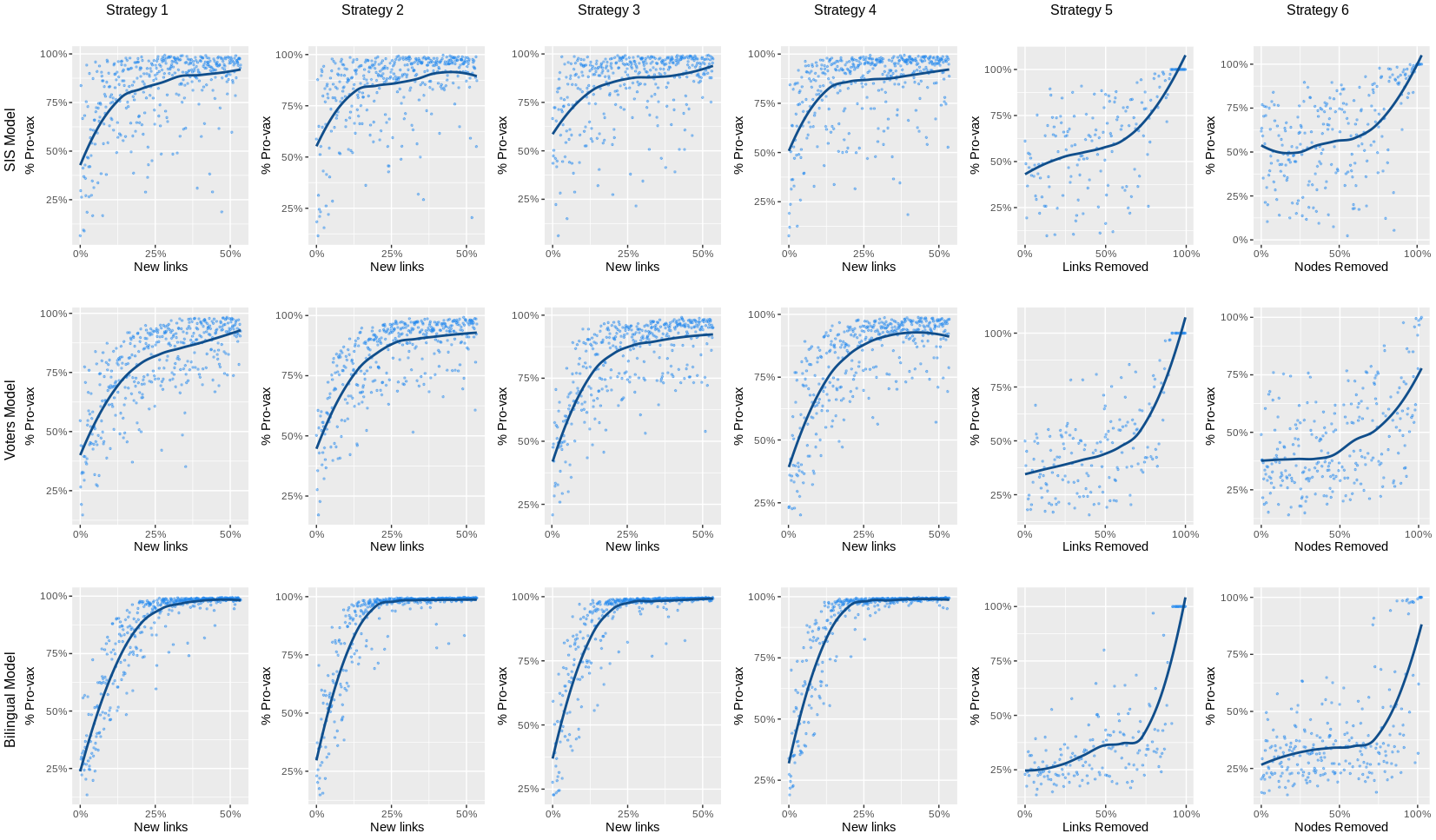}
\caption{The consequences of the strategies are shown in all the models. Each point represents a simulation, on x-axis the number of links (or nodes) added (or removed), on the y axis the proportion of undecided users attracted by pro-vax community.
A LOESS curve is depicted to resume the effect of the increase of the intervention. The parameters of the models are set $\beta$ = 0.4, $\mu$ = 0.2, $\gamma$ = 0.1, \emph{d} = 0.1. }
\label{all_strategies}
\end{figure}
In Fig. ~\ref{all_strategies} the effects of the strategies are compared. \\
\begin{itemize}
    \item The first aspect that has to be highlighted is the great difference between linkage and censoring strategies: in all the cases, it is enough to increase the number of links by 16$\%$ to obtain at least 75$\%$ of consensus of neutral population. A small linkage intervention on the network provides satisfying results. On the other hand, a huge censoring intervention is needed to effectively obstruct no-vax narrative. The removal of 50$\%$ of no-vax pages or connections is not enough to prevent no-vax overcome in Bilingual and Voters Models.
    \item It is possible to observe a different behaviour of Bilingual Model. In this case, we recall that neutral nodes tend to place only at the border of the communities, leading to the strongest polarisation among our models: for this reason, communities assume homogeneous colouring and it is almost impossible to  have mixed anti-/pro-vaccines clusters. When one party is able to conquer one community, all the nodes choose the same polarity. Thus, tie configurations are not allowed: small interventions can completely change the fate, from a no-vax consensus to a pro-vax consensus, without passing through intermediate steps.
    \item Linkage strategies have very similar results: it is not necessary to point strategic nodes, the influence of pro-vax community can grow just by increasing the connections between pro-vax pages and the rest of the network.
\end{itemize}
The initial problem is that pro-vax pages are too far from the centre of the network, and undecided cluster tend to follow no-vax opinion, that is infiltrated at the core of the network. Removing some nodes or links the problem persists. Even if some of them are censored, the opportunities to be in contact with their narratives are everywhere.\\
Moreover, we have to consider that policies aimed to stop all the opposing opinions are not feasible. \\
Based on our results, widening the pro-vax audience could be effective to prevent no-vax spread in online social network.

\section*{Conclusions}
The models analysed describe different dynamics that can occur during a news spread related to vaccination debate. Neutral cluster can represent indifferent population (SIS Model, they are not interested on the topic), centrist party (Voters Model, they propose an argument different from the other factions), or compromise option (Bilingual Model, they agree with both the positions). The models lead to different levels of polarisation and no-vax success: however, under each setting, the no-vax community is able to attract a large part of undecided community, growing much more than pro-vax party. Even if no-vax community is much less populated than pro-vax one, their strategic disposition provides worthwhile results.\\
Pro-vax pages are too far from the core of the network, tending to sustain each other without involving neutral cluster in the discussion. On the other hand, no-vax party offers many points of engagement to neutrals: having many small pages sparse in the network, they can propose many different narratives to persuade neutral population.\\
Under SIS Model, neutral nodes change side many times, and, hearing the opinion from many sources, it is more likely that they follow pro-vax opinion. The other two models increase the effect of the initial configuration: it is more likely to follow the first narrative encountered. This favours no-vax faction, that create a strong region at the centre of the network that survives over the simulations, leading to no-vax victory. \\
Exploring the effects of our interventions, we have seen that it is possible to contrast this fate. In fact, a small random increasing of  the connections between pro-vax pages and the rest of the network provides a persuasion of 90$\%$ of undecided population. It is not needed to target strategic nodes, since Random Linkage Strategy is very effective. On the other side, our simulations tell us that censor is absolutely ineffectual: even if we remove 50$\%$ of no-vax contents, neutral pages tend to side anti-vaccines part. Indeed, we can weaken no-vax community, but pro-vax pages remain too peripheral, without attacking the central battlefield.\\
Thus, the only feasible way to obstruct anti-scientific narratives is to make the population aware of authoritative and credible opinions, and our efforts should focus on the widening the audience of trustworthy sources. Our study lead to think that scientific view is too elitist: high credible narratives are too closed in their own community, failing in communicating their messages to most of the population.\\
Our study is limited to Facebook pages data, it would be interesting to compare the models outcomes on different data, coming from other social networks or relating to other polarised debates with a consistent undecided population (such as politics or climate change).\\
Future studies could be aimed to recognise the nodes that maximise the influence in our models: which pages are most effective in persuading new undecided nodes, and which neutral pages should be the main target for extreme factions. \\
Monitoring our network over the years, it would be interesting to accurately model the evolution of the pages and the link prediction, to predict the effect of the news spread in future times. 
\section*{Limits of the Model}
It is fair to consider all the limits of the models and the data presented in this paper. The models are not thought to predict the real behaviour of the social network, but to catch and replicate some dynamics that emerge in an opinion dynamics context.\\
Our data refers to Facebook pages in 2019: to have a complete view, it is needed to compare the results with other time periods or to different social networks, since the interactions could have significant differences.\\
We also have to point out that it is not possible to describe human behaviour from social networks data: Facebook (or Twitter, Reddit, etc.) is just a limited insight of the society, where people can be pushed to share their own opinions (to state an agreement or to feel popular).

\newpage
\bibliographystyle{plain}
\bibliography{NovaxBib} 

\begin{thebibliography}{10}

\bibitem{pandora}
Kata A.
\newblock A postmodern pandora's box: Anti-vaccination misinformation on the
  internet.
\newblock {\em Vaccine}, page 28(7), 2010.

\bibitem{detecting}
Gema Bello~Orgaz, Julio Hernandez-Castro, and David Camacho.
\newblock Detecting discussion communities on vaccination in twitter.
\newblock {\em Future Generation Computer System}, page~66, 2016.

\bibitem{opinionepidem}
Luís Bettencourt, Ariel Cintrón-Arias, David Kaiser, and Carlos
  Castillo-Chávez.
\newblock The power of a good idea: Quantitative modeling of the spread of
  ideas from epidemiological models.
\newblock {\em Phys. A}, pages 364, 513–536, 2006.

\bibitem{opinionstats}
Claudio Castellano, Santo Fortunato, and Vittorio Loreto.
\newblock Statistical physics of social dynamics.
\newblock {\em Rev. Mod. Phys.}, pages 81(2), 591, 2009.

\bibitem{opinioncrime}
Rafael Curiel and Steven Bishop.
\newblock Modelling the fear of crime.
\newblock {\em Proc. R. Soc. Lond. A Math. Phys. Eng. Sci}, page 2017, 473,
  2203.

\bibitem{opinionpersuasion}
John~P. Curtis and Frank~T. Smith.
\newblock The dynamics of persuasion.
\newblock {\em Int. J. Math. Models Methods Appl. Sci.}, pages 2(1), 115–122,
  2008.

\bibitem{polarizationandfakenews}
Michela Del~Vicario, Walter Quattrociocchi, Antonio Scala, and Fabiana Zollo.
\newblock Polarization and fake news: Early warning of potential misinformation
  targets.
\newblock {\em ACM Transactions on the Web}, page~13, 2019.

\bibitem{opinionleader}
Bertram Düring, Peter Markowich, Jan-Frederik Pietschmann, and Marie-Therese
  Wolfram.
\newblock Boltzmann and fokker–planck equations modelling opinion formation
  in the presence of strong leaders.
\newblock {\em Proc. R. Soc. Lond. A Math. Phys. Eng. Sci.}, pages 465(2112),
  3687–3708, 2009.

\bibitem{opinionsegr}
Bertram Düring and Marie-Therese Wolfram.
\newblock Opinion dynamics: Inhomogeneous boltzmann-type equations modelling
  opinion leadership and political segregation.
\newblock {\em Proc. R. Soc. Lond. A Math. Phys. Eng. Sci.}, pages 471, 2182,
  2015.

\bibitem{orderingbilingual}
Víctor Eguíluz and Maxi Miguel.
\newblock Ordering dynamics with two non-excluding options: Bilingualism in
  language competition.
\newblock {\em New Journal of Physics}, page 8(308), 2006.

\bibitem{opinionkine}
Toscani G.
\newblock Kinetic models of opinion formation.
\newblock {\em Commun. Math. Sci.}, pages 4(3), 481–496, 2006.

\bibitem{pinterest}
Jeanine Guidry, Kellie Carlyle, Marcus Messner, and Yan Jin.
\newblock On pins and needles: How vaccines are portrayed on pinterest.
\newblock {\em Vaccine}, page 33(39), 2015.

\bibitem{competition}
Neil Johnson, Nicolas Velasquez, Nicholas Restrepo, Rhys Leahy, Nicholas
  Gabriel, Sara Oud, Minzhang Zheng, Pedro Manrique, Stefan Wuchty, and Yonatan
  Lupu.
\newblock The online competition between pro- and anti-vaccination views.
\newblock {\em Nature}, page 582, 2020.

\bibitem{vaccineautism}
Mo~Jones~Jang, Brooke Mckeever, Robert Mckeever, and Joon Kim.
\newblock From social media to mainstream news: The information flow of the
  vaccine-autism controversy in the us, canada, and the uk.
\newblock {\em Health Communication}, pages 34:1, 110--117, 2019.

\bibitem{evencovid}
Megget K.
\newblock Even covid-19 can't kill the anti-vaccination movement.
\newblock {\em BMJ}, page 369, 2020.

\bibitem{trust}
Carl Latkin, Lauren Dayton, Grace Yi, Arianna Konstantopoulos, and Basmattee
  Boodram.
\newblock Trust in a covid-19 vaccine in the u.s.: A social-ecological
  perspective.
\newblock {\em Social Science and Medicine}, page 270, 2021.

\bibitem{databasecovid}
Edouard Mathieu, Hannah Ritchie, Esteban Ortiz-Ospina, Max Roser, Joe Hasell,
  and Charlie Giattino.
\newblock A global database of covid-19 vaccinations.
\newblock {\em Nat Hum Behav}, page 5(7), 2021.

\bibitem{measlesdata1}
Government of~Philippines.
\newblock Ndrrmc update sitrep no. 13 re measles outbreak.
\newblock {\em National Disaster \& Risk Reduction and Management Council},
  2019.

\bibitem{measlesdata2}
Government of~Samoa.
\newblock Population demography indicator summary.
\newblock {\em Samoa Bureau of Statistics}, 2019.

\bibitem{efforts}
Ball P.
\newblock Anti-vaccines movement might undermine pandemic efforts.
\newblock {\em Nature}, pages 581,251, 2020.

\bibitem{susceptibility}
Jon Roozenbeek, Claudia Schneider, Sarah Dryhurst, John Kerr, Alexandra
  Freeman, Gabriel Recchia, Anne~Marthe van~der Bles, and Sander van~der
  Linden.
\newblock Susceptibility to misinformation about covid-19 around the world.
\newblock {\em Royal Society Open Science}, page 7(10), 2020.

\bibitem{factor}
Daniel Salmon, Lawrence Moulton, Saad Omer, M.~Dehart, Shannon Stokley, and
  Neal Halsey.
\newblock Factors associated with refusal of childhood vaccines among parents
  of school-aged children: a case-control study.
\newblock {\em Archives of Pediatrics and Adolescent Medicine}, page 159(5),
  2005.

\bibitem{polarization}
Ana Schmidt, Fabiana Zollo, Antonio Scala, Cornelia Betsch, and Walter
  Quattrociocchi.
\newblock Polarization of the vaccination debate on facebook.
\newblock {\em Vaccine}, pages 36(25), 3606--3612, 2018.

\bibitem{understanding}
Brit Trogen and Liise-Anne Pirofski.
\newblock Understanding vaccine hesitancy in covid-19.
\newblock {\em Med N Y}, pages 2(5), 498–501, 2021.

\bibitem{freezing}
Federico Vazquez, Pavel Krapivsky, and Sidney Redner.
\newblock Constrained opinion dynamics: freezing and slow evolution.
\newblock {\em J. Phys. A: Math. Gen.}, page~36, 2003.

\bibitem{ultimatefate}
Federico Vazquez and Sidney Redner.
\newblock Ultimate fate of constrained voters.
\newblock {\em J. Phys. A: Gen. Phys.}, page~37, 2004.

\bibitem{vosoughi18}
S.~Vosoughi, D.~Roy, and S.~Aral.
\newblock The spread of true and false news online.
\newblock {\em Science}, pages 1146--1151, mar 2018.

\bibitem{socmediavachesitancy}
Steven Wilson and Charles Wiysonge.
\newblock Social media and vaccine hesitancy.
\newblock {\em BMJ Global Health}, page~5, 2020.

\end{thebibliography}







\end{document}